\RequirePackage{silence}
\documentclass[twocolumn]{aast}
\newcommand{\lv}{1.08}

\newcommand{\blue}[1]{{#1}}
\definecolor{TealBlue}{rgb}{0.0, 0.5, 0.5}

\newcommand{\devcol}{\mc{1}{c}{Dev.}}
\newcommand{\unccol}{\mc{1}{c}{Unc.}}
\AtBeginDocument{%
  \let\oldsection\section
  \renewcommand{\section}[1]{\oldsection{\small #1}}%
}

\usepackage{capt-of}
\usepackage{hyperref}

\newcommand{\ie}{{\it i.e.}}

\AtBeginEnvironment{tablenotes}{\footnotesize}

\RequirePackage{etoolbox}

\newcommand{\defineSpecies}[2]{\csdef{spec@#1}{#2}}
\newrobustcmd{\spec}[1]{%
  \ifcsname spec@#1\endcsname%
    \csuse{spec@#1}%
  \else
    \GenericError{}{Undefined species `#1'}{}{}
  \fi
}

\RequirePackage{float}
\RequirePackage{longtable}
\RequirePackage{booktabs}
\RequirePackage{threeparttable}
\RequirePackage{physics}
\RequirePackage{amsmath}
\RequirePackage{bm}
\RequirePackage{array}
\RequirePackage{multirow}
\RequirePackage{chngcntr}
\RequirePackage{enumerate}
\RequirePackage{upgreek}
\RequirePackage{changepage}

\usepackage{setspace}

\usepackage{silence}

\usepackage{array}
\usepackage{silence}
\defineSpecies{h216o}{H$_2^{~16}$O}
\defineSpecies{h2s}{H$_2$S}
\defineSpecies{h217o}{H$_2^{~17}$O}
\defineSpecies{h218o}{H$_2^{~18}$O}
\defineSpecies{h2o}{H$_2$O}
\defineSpecies{d216o}{D$_2^{~16}$O}
\defineSpecies{d217o}{D$_2^{~17}$O}
\defineSpecies{d218o}{D$_2^{~18}$O}
\defineSpecies{d2o}{D$_2$O}
\defineSpecies{hd16o}{HD$^{16}$O}
\defineSpecies{hd17o}{HD$^{17}$O}
\defineSpecies{hd18o}{HD$^{18}$O}
\defineSpecies{hdo}{HDO}
\defineSpecies{12c2}{$^{12}$C$_2$}
\defineSpecies{14nh3}{$^{14}$NH$_3$}
\defineSpecies{48ti16o}{$^{48}$Ti$^{16}$O}
\defineSpecies{h232s}{H$_2^{~32}$S}
\defineSpecies{so2}{SO$_2$}
\defineSpecies{90zr16o}{$^{90}$Zr$^{16}$O}
\defineSpecies{12c2h2}{$^{12}$C$_2$H$_2$}
\defineSpecies{h3+}{H$_3^+$}
\defineSpecies{h2d+}{H$_2$D$^+$}
\defineSpecies{d2h+}{D$_2$H$^+$}
\defineSpecies{16o}{$^{16}$O}
\defineSpecies{17o}{$^{17}$O}
\defineSpecies{18o}{$^{18}$O}
\defineSpecies{arno+}{Ar$\cdot$NO$^{+}$}
\defineSpecies{16o3}{$^{16}$O$_3$}
\defineSpecies{nh3}{NH$_3$}
\defineSpecies{13cs2}{$^{13}$CS$_2$}
\defineSpecies{co2}{CO$_2$}
\defineSpecies{ch3f}{CH$_3$F}
\defineSpecies{n2o}{N$_2$O}
\defineSpecies{co}{CO}
\defineSpecies{ch4}{CH$_4$}
\defineSpecies{n2}{N$_2$}
\defineSpecies{h2}{H$_2$}
\defineSpecies{h2xo}{H$_2^{~X}$O}
\defineSpecies{hdxo}{HD$^X$O}
\defineSpecies{d2xo}{D$_2^{~X}$O}
\defineSpecies{o2}{O$_2$}
\defineSpecies{o3}{O$_3$}
\defineSpecies{ch4:ch4}{CH$_4$:CH$_4$}
\defineSpecies{c2h2}{C$_2$H$_2$}
\defineSpecies{nh2d}{NH$_2$D}

\newrobustcmd{\param}[1]{%
  \ifcsname param@#1\endcsname
    \csuse{param@#1}
  \else
    \GenericError{}{Undefined param `#1'}{}{}
  \fi
}

\newcommand{\const}[1]{#1}
\newcommand{\mc}{\multicolumn}
\newcommand{\mr}{\multirow}

\newcommand{\khzipa}{kHz\,Pa$^{-1}$}

\newcommand{\PreserveBackslash}[1]{\let\temp=\\#1\let\\=\temp}
\newcolumntype{C}[1]{>{\PreserveBackslash\centering}p{#1}}
\newcolumntype{R}[1]{>{\PreserveBackslash\raggedleft}p{#1}}
\newcolumntype{L}[1]{>{\PreserveBackslash\raggedright}p{#1}}

\newcommand{\ttcap}{}

\newcommand{\tcap}{}

\newcommand{\tlab}{}

\begin{document}

\title{\vspace{-0.8cm}\Large Precise frequencies of \spec{h216o} 
lines protected for radio astronomy}

\correspondingauthor{Wim Ubachs \& Roland Tóbiás}
\email{w.m.g.ubachs@vu.nl \& roland.tobias@ttk.elte.hu}

\author[0009-0000-3924-3497]{Arthémise Altman}
\affiliation{Institute of Condensed Matter and Nanosciences, 
Université catholique de Louvain, Louvain-la-Neuve, Belgium}

\author[0000-0003-3674-5066]{Roland Tóbiás}
\affiliation{Department of Chemistry, University of Vermont, Burlington, 
          VT, 05405, USA \& \\ 
          Institute  of Chemistry, ELTE E\"otv\"os Lor\'and University,
          H-1117 Budapest, P\'azm\'any P\'eter s\'et\'any 1/A, Hungary}

\author[0000-0002-6675-0969]{Alexandr S. Bogomolov}
\affiliation{Institute of Condensed Matter and Nanosciences, 
Université catholique de Louvain, Louvain-la-Neuve, Belgium}

\author[0000-0003-1263-6003]{Meissa L. Diouf}
\affiliation{Department of Physics and Astronomy, LaserLaB, 
Vrije Universiteit, De Boelelaan 1100, 1081 HZ Amsterdam, 
The Netherlands}

\author[0000-0002-8169-5961]{Frank M. J. Cozijn}
\affiliation{Department of Physics and Astronomy, LaserLaB, 
Vrije Universiteit, De Boelelaan 1100, 1081 HZ Amsterdam, 
The Netherlands}

\author[0000-0001-3674-191X]{Attila G. Cs\'asz\'ar}
\affiliation{Institute of Chemistry, ELTE Eötvös Loránd
University, H-1518 Budapest, P.O. Box 32, Hungary}

\author[0000-0001-5044-5137]{Clément Lauzin}
\affiliation{Institute of Condensed Matter and Nanosciences, 
Université catholique de Louvain, Louvain-la-Neuve, Belgium}

\author[0000-0001-7840-3756]{Wim Ubachs}
\affiliation{Department of Physics and Astronomy, LaserLaB, 
Vrije Universiteit, De Boelelaan 1100, 1081 HZ Amsterdam, 
The Netherlands}

\begin{abstract}
Precise frequency values have been determined for \spec{h216o} radio lines appearing 
in protected line lists of the International Astronomical Union and the Panel on 
Frequency Allocations of the US National Academy of Sciences.
The improved precision is attributable to a spectroscopic network built from a large 
set of near-infrared Lamb-dip lines augmented with a handful of ultrahigh-accuracy 
rotational transitions.
The ultraprecise \spec{h216o} network contains \const{376} Lamb-dip lines recorded
previously \emph{via} our frequency-comb locked cavity-enhanced spectrometer.
During the present study, altogether \const{55} target lines have been (re)measured 
at high accuracy.
Due to our network-assisted measurements, the accuracy has been significantly 
improved with respect to previous direct radio-frequency measurements for all the 
protected lines of \spec{h216o} between 750 and 3000 GHz. 
Furthermore, \const{43} of these protected transitions now benefit from the accuracy
of the new near-infrared Lamb dips reported in this paper.
\end{abstract}

\section{Introduction}
\label{sec:intro}

The water molecule, \spec{h2o}, is one of the key species throughout the universe.
It is made up of hydrogen,
the most abundant element produced primordially, and oxygen, 
the third most abundant element produced in supernovae~\citep{13DiHeNe}.
Water molecules are synthesized in either gas-phase interactions~\citep{72SoKl}
or \emph{via} solid-state processes~\citep{82TiHa} in interstellar space.

Water in the interstellar medium was discovered in the early days of radio 
astronomy~\citep{69ChRaToTh} through a 22 GHz maser line.
Since then, the presence of water has been investigated \emph{via} radio 
astronomical observations of the interstellar medium under various conditions
\citep{00WrDiBlFe,06CeGoDaLe}, using maser transitions as well as
thermal spectral lines.
In pushing the limits of radio astronomy, water was observed at a redshift of
$z=5.6$~\citep{13WeBrMaVi}.

Datasets for reference frequencies of many molecular transitions, including those of 
\spec{h216o}, are maintained in numerous databases.
In terms of radio astronomy, the most relevant databanks appear to be the Cologne 
Database for Molecular Spectroscopy
(CDMS)\footnote{\url{https://cdms.astro.uni-koeln.de/}}~\citep{05MuScStWi,16ChScScSt} 
and the Jet Propulsion Laboratory (JPL)
line catalog\footnote{\url{https://spec.jpl.nasa.gov/}} \citep{98PiPoCoDe}.

At the XXIst General Assembly of the International Astronomical Union (IAU),
taking place in Buenos Aires (July 23 -- August 1, 1991),
the astronomically most important spectral lines have been carefully reviewed.
The need to protect the frequencies of these lines with some minimally 
line-specific band widths was defined. 
The original list covered 73 lines, of which four pertain to \spec{h216o}.
Later, in preparation for the World Radiocommunication Conference 2000, held 
in Istanbul (May 8 -- June 2, 2000), the IUCAF (Scientific Committee on Frequency 
Allocations for Radio Astronomy and Space Science)
mm-wavelengths Working Group examined all known transitions in this region and 
specified revised allocations above 71 GHz. 
This catalog supplements the earlier list constructed by the IAU, extends to 1 THz, 
and covers in total 13 lines for \spec{h216o}.
Both collections can be accessed
digitally.\footnote{www.craf.eu/iau-list-of-important-spectral-lines/}
Fifteen years later, the Panel on Frequency Allocations and
Spectrum Protection for Scientific Uses of the US National Academy of
Sciences \citep{15Committe} extended the list of protected water lines, in particular 
covering the frequency window between 1 and 3 THz, with 47 lines for \spec{h216o} 
(here referred to as the NAS list).
This leads to altogether \const{60} protected lines for the main
isotopologue of water.

The principal aim of this investigation has been to provide frequency estimates with
improved accuracy for the overwhelming majority of these \const{60} lines.
To achieve this goal, laser-based precision-spectroscopy measurements
were conducted for dozens of carefully chosen near-infrared transitions
in the 1.2\,--\,1.4 ${\upmu}$m range.
Based on all the new and old kHz-accuracy transitions available for \spec{h216o}, 
an ultraprecise spectroscopic network \citep{11CsFu} was assembled,
delivering highly accurate energies for 79 pure rotational states.

As shown in \cite{24UbCsDiCo}, such ultraprecise spectroscopic networks facilitate
the  extraction of highly accurate frequencies for radio lines.
In that study, the existing ultraprecise \spec{h216o} and \spec{h218o} networks 
were extended with newly measured Lamb-dip lines, enabling the derivation of
accurate frequencies for a list of \spec{h216o} maser transitions, as well as 
\spec{h218o} lines observed extra-galactically.
Following this approach, the present work provides significantly refined 
frequency predictions for the protected \spec{h216o} transitions, enhancing their 
reliability for future astronomical applications.
\blue{The precise frequencies may find application in the accurate determination of 
velocities in astrophysical environments \citep{04CaPuLa}.}

\section{Ultraprecise H$_2^{~16}$O network}\label{sec:network}
As to the notation used for the rovibrational states of \spec{h216o}, 
the vibrations are labelled with a triplet of normal-mode quantum numbers, 
$(v_1\,v_2\,v_3)$, where $v_1$, $v_2$, and $v_3$ represent the symmetric stretch,
bend, and antisymmetric stretch motions, respectively.
For the rotations, the rigid-rotor asymmetric-top quantum numbers $J$, $K_a$, and 
$K_c$ are employed, in the form $J_{K_a,K_c}$.
The rovibrational assignment of a transition is given here as $(v'_1\,v'_2\,v'_3)
J'_{K'_a,K'_c} 
\leftarrow (v''_1\,v''_2\,v''_3)J''_{K''_a,K''_c}$ , where $'$ and $''$ stand for 
its upper and lower states, respectively.
Throughout the whole paper, only standard \citep{nistunc} uncertainties will be 
used.

In this study, the spectroscopic-network-assisted precision spectroscopy (SNAPS)
procedure \citep{20ToFuSiCs} was applied to determine a large number of
accurate transition frequencies, involving newly measured near-infrared 
Lamb-dip lines to form energy differences.
These Lamb dips were recorded \emph{via} laser-based Doppler-free saturation 
spectroscopy, yielding frequency uncertainties down to the kHz level. 
The superior sensitivity of the noise-immune cavity-enhanced 
optical-heterodyne molecular spectroscopy (NICE-OHMS) method \citep{24GiHuUb}
is well-suited to detect weak lines, whose lower states have small 
population densities at room temperature.
When combined with frequency-comb calibration, NICE-OHMS is 
suitable for providing absolute accuracy for the measured 
line positions.

In our previous experimental campaigns~\citep{20ToFuSiCs,22DiToScCo,
24ToDiCoUb,24UbCsDiCo}, several hundred \spec{h216o} Lamb dips were measured, 
involving rovibrational states within
polyads $P= 0$, 1, 4, and 5, where $P=2v_1 +v_2 + 2v_3$ is the polyad number.
Note, in particular, that ultrahigh-accuracy energies could be deduced for
so-called hubs (that is, quantum states holding the largest number of transitions) 
in the full experimental \spec{h216o} network \citep{20FuToTePo,20FuToTePob}
available at that time \citep{24ToDiCoUb}.

From the perspective of rovibrational spectroscopy, the quantum states of the 
\spec{h216o} species, containing two equivalent hydrogen atoms with nuclear 
spin of $I=1/2$, fall into four subnetworks.
The \textit{para}-\spec{h216o} levels, with a total nuclear spin of $I=0$,
are fully separated from the \textit{ortho}-\spec{h216o} states, with $I=1$,
and no transitions between the states of these nuclear-spin isomers
have ever been detected~\citep{04MiTe}.
The \textit{para}- and \textit{ortho}-\spec{h216o} levels are defined by
$(-1)^{v_3+K_a+K_c}$ being $+1$ or $-1$, respectively.
Within the $(0\,0\,0)$ vibrational state, levels of opposite (total) 
parity cannot be linked \emph{via} near-infrared dipole lines
(note that \emph{even}/\emph{odd}-parity states have even/odd $K_c$ values).
Thus, when building a network from them, the even- and odd-parity
subnetworks must be connected by pure rotational transitions. 

One of the key features of spectroscopic networks is that their transitions
form so-called paths and cycles (\ie, series of connected lines, where 
the starting and ending states are different and identical, respectively).
Paths are highly useful as they define energy differences between their 
starting and ending states, along with conservative uncertainty estimates.
Cycles help not only to confirm the internal accuracy of the measurements,
but also to check the self-consistency of the assignments.
For further details, see Sec. 4 of \citet{24UbCsDiCo}.

In the ultraprecise network built for \spec{h216o}, several microwave
\citep{69Kukolich,06GoMaGuKn,09CaPuHaGa} and near-infrared \citep{18KaStCaDa,
18ChHuTaSu,20ToFuSiCs,22KaLaChCa,22DiToScCo,24ToDiCoUb,24UbCsDiCo}
transitions have been included. 
Following its latest update \citep{24UbCsDiCo}, this network now contains 
411 previously measured microwave
and near-infrared lines, where all the \emph{ortho} (\emph{para}) transitions
form a single \emph{ortho} (\emph{para}) component.

\section{NICE-OHMS measurement campaigns} \label{sec:exp}

To deduce kHz-accuracy frequencies for the protected \spec{h216o} lines, a well-%
selected set of near-infrared lines was recorded \emph{via} our NICE-OHMS setup at 
room temperature.
The first part of the experiments was performed at LaserLaB Amsterdam, with the 
NICE-OHMS spectrometer developed there~\citep{18CoDuSaEi,20ToFuSiCs}.
This setup was later moved to Louvain-la-Neuve, where the UCLouvain laboratory
served as the site for the second set of measurements.

\begin{figure}[b]
\centering
\hspace{-0.2cm}\includegraphics[width=0.95\linewidth]{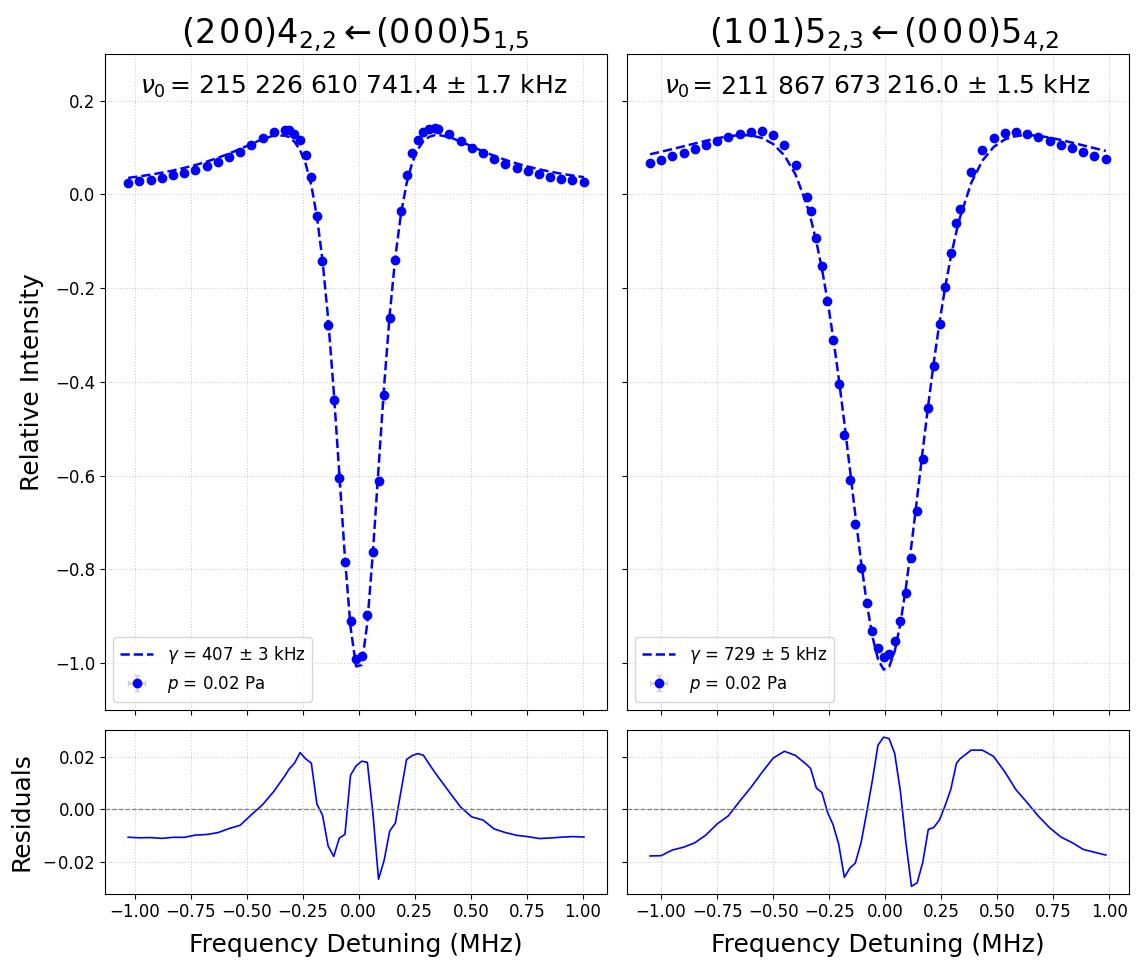}
\caption{\hspace{-0.2cm} Examples for rovibrational Lamb-dip spectra measured 
         for \spec{h216o} \emph{via} our NICE-OHMS setup. 
         \normalfont{The $\gamma$ parameters represent the full width at half 
         maximum (FWHM) of the fitted Lorentzians.
         The frequency detuning is relative to the observed 
         frequency, $\nu_0$ (see also Table~\ref{table:meas}).}
         }
\label{fig:spectra}
\vspace{-0.0cm}
\end{figure}

In the Amsterdam campaign, statistical averaging was carried out over four 
recordings per line. 
The frequency metrology was based on beat-note measurements against a frequency-%
comb laser (Menlo Systems, FC1500), stabilized to a Cs clock (Microsemi CSIII),
and long-term disciplined with respect to GPS.
This yielded a frequency scale with negligible uncertainties compared to the
statistical errors.
The pressure-shift uncertainties were estimated using a worst-case pressure slope,
found to be $\pm$20 \khzipa\ in our previous SNAPS studies \citep{20ToFuSiCs,
22DiToScCo}.
Since the measurements were carried out at 0.01--0.02 Pa, this led to a 
pressure-induced uncertainty of 0.2--0.4 kHz.
In the final uncertainties, the power-shift and day-to-day errors (0.5 and 1.5 
kHz, respectively) were also taken into account.

For the measurement campaign at Louvain-la-Neuve, an automated scanning procedure 
was implemented, ensuring multiple recordings under stabilized conditions. 
All these experiments were conducted at 0.02 Pa, averaging over at least 16 scans,
which gave rise to deviations typically below 1 kHz.
At Louvain-la-Neuve, the frequency-comb laser (Toptica DFC Core+) was locked to a
quartz oscillator clock, whose uncertainty was studied in a separate measurement 
campaign.
This contribution is 0.9 kHz for the majority of lines averaged over 16 
recordings, a somewhat larger value for averages over 8 scans, while  for the four 
lines involved in long-term averaging over 100 scans this clock-related uncertainty 
becomes negligible. 
In the total uncertainty budget, the pressure-shift uncertainty, translating to
0.4 kHz at 0.02 Pa, was added as a linear term to the combined statistical (clock 
plus scan) uncertainty.

Figure~\ref{fig:spectra} shows spectra of two new Lamb dips recorded 
at Louvain-la-Neuve and fitted using a derivative dispersion function
(this model is well suited to wavelength-modulated NICE-OHMS signals).
On the bottom panels of Fig.~\ref{fig:spectra}, the fitting residuals
are also plotted, whose oscillating features are ascribed to effects
due to the combination of \blue{high-frequency modulation with low frequency 
wavelength modulation, a phenomenon investigated in detail by 
\citet{Supplee1994}.
These oscillating features are characteristic of using a large modulation 
depth, where the signal deviates from a pure first-derivative line shape and 
contains contributions from higher-order derivatives of the dispersion profile. 
While this complicates a fully analytical description, the symmetry of these 
contributions ensures that their effect on the determination of the line 
center frequency is negligible.
Note that the verification cycles, presented in Table S3, 
also demonstrate that the new lines are consistent within the 
claimed uncertainties.
}

\linespread{1.5}
\begin{widetext}
\begingroup
\linespread{0.9}
\vspace{-0.3cm}
\setlength{\tabcolsep}{5pt} 

\renewcommand{\ttcap}{}
\renewcommand{\tcap}{New near-infrared Lamb-dip lines measured \emph{via} an ultrasensitive 
          NICE-OHMS spectrometer for \spec{h216o}.}
\renewcommand{\tlab}{table:meas}
\begingroup
\scriptsize
\renewcommand{\u}{\underline}
\begin{longtable}{rccrrl}
\caption[\ttcap]{\tcap} \\[-10pt]
\label{\tlab} \\
\hline \hline \\[-10pt]
\endfirsthead
\mc{6}{c}{\tablename\ \thetable\ -- \emph{Continued from previous page}}\\ 
\hline
\mc{2}{c}{NICE-OHMS (this work)$^a$}       & \mr{2}{*}{Rovibrational assignment$^b$}                     & \mc{3}{c}{Previous results$^c$}                 \\
\cmidrule{1-2}\cmidrule{4-6} \\[-12pt]
                    Line frequency\,/\,kHz & $p$\,/\,Pa &                                                &      \devcol &      \unccol & \mc{1}{c}{Reference}\\[2pt] \hline
\endhead
\hline \mc{6}{r}{\emph{ Continued on next page }} \\
\endfoot
\hline \hline
\endlastfoot
\mc{2}{c}{NICE-OHMS (this work)$^a$}       & \mr{2}{*}{Rovibrational assignment$^b$}                     & \mc{3}{c}{Previous results$^c$/kHz}                 \\
\cmidrule{1-2}\cmidrule{4-6} \\[-12pt]
                    Line frequency\,/\,kHz & $p$\,/\,Pa &                                                &      \devcol      &      \unccol & \mc{1}{c}{Reference} \\[2pt] \hline
            $209\,966\,584\,532.4 \pm 1.4$ &       0.02 & $(1\,0\,1)3_{1,3} \leftarrow (0\,0\,0)4_{3,2}$ &         --300~~~~ &     3000~~~~ & \cite{18MiMoKaKa}    \\
            $210\,038\,636\,486.7 \pm 1.4$ &       0.02 & $(2\,0\,0)5_{0,5} \leftarrow (0\,0\,0)5_{3,2}$ &         --3.7~~~~ &     10.4~~~~ & \cite{24ToDiCoUb}    \\
            $210\,463\,484\,645.5 \pm 1.5$ &       0.02 & $(0\,2\,1)6_{3,4} \leftarrow (0\,0\,0)6_{1,5}$ &           500~~~~ &     3000~~~~ & \cite{18MiMoKaKa}    \\
            $210\,664\,080\,420.2 \pm 1.3$ &       0.02 & $(0\,2\,1)6_{5,1} \leftarrow (0\,0\,0)7_{3,4}$ & \u{--21\,800}~~~~ &     3000~~~~ & \cite{18MiMoKaKa}    \\
            $211\,120\,145\,834.0 \pm 1.5$ &       0.02 & $(2\,0\,0)5_{4,2} \leftarrow (0\,0\,0)5_{5,1}$ &         --5.2~~~~ &      3.0~~~~ & \cite{24UbCsDiCo}    \\
            $211\,126\,350\,475.9 \pm 1.5$ &       0.02 & $(2\,0\,0)5_{4,1} \leftarrow (0\,0\,0)5_{5,0}$ &         --0.3~~~~ &     11.1~~~~ & \cite{24ToDiCoUb}    \\
            $211\,545\,072\,422.8 \pm 1.5$ &       0.02 & $(1\,0\,1)2_{1,1} \leftarrow (0\,0\,0)3_{3,0}$ &         --0.8~~~~ &      7.5~~~~ & \cite{24ToDiCoUb}    \\
            $211\,630\,806\,780.1 \pm 0.6$ &       0.02 & $(0\,2\,1)6_{3,4} \leftarrow (0\,0\,0)5_{3,3}$ &           100~~~~ &     3000~~~~ & \cite{18MiMoKaKa}    \\
            $211\,802\,311\,374.9 \pm 1.4$ &       0.02 & $(2\,0\,0)6_{3,4} \leftarrow (0\,0\,0)6_{4,3}$ &         --9.6~~~~ &     10.0~~~~ & \cite{24ToDiCoUb}    \\
            $211\,867\,673\,216.0 \pm 1.5$ &       0.02 & $(1\,0\,1)5_{2,3} \leftarrow (0\,0\,0)5_{4,2}$ &         --0.2~~~~ &      7.7~~~~ & \cite{24ToDiCoUb}    \\
            $212\,286\,960\,311.8 \pm 1.4$ &       0.02 & $(2\,0\,0)6_{3,3} \leftarrow (0\,0\,0)6_{4,2}$ &           2.5~~~~ &     10.4~~~~ & \cite{24ToDiCoUb}    \\
            $212\,549\,726\,956.5 \pm 1.5$ &       0.02 & $(2\,0\,0)4_{2,3} \leftarrow (0\,0\,0)5_{1,4}$ &           1.3~~~~ &      7.5~~~~ & \cite{22DiToScCo}    \\
            $212\,843\,780\,340.2 \pm 1.4$ &       0.02 & $(2\,0\,0)5_{1,5} \leftarrow (0\,0\,0)5_{2,4}$ &           3.2~~~~ &      7.6~~~~ & \cite{22DiToScCo}    \\
            $212\,877\,518\,672.9 \pm 1.8$ &       0.02 & $(1\,0\,1)3_{1,3} \leftarrow (0\,0\,0)3_{3,0}$ &           300~~~~ &     3000~~~~ & \cite{18MiMoKaKa}    \\
            $213\,248\,005\,986.5 \pm 1.7$ &       0.02 & $(2\,0\,0)3_{2,2} \leftarrow (0\,0\,0)3_{3,1}$ &           1.5~~~~ &      5.2~~~~ & \cite{20ToFuSiCs}    \\
            $213\,317\,003\,150.6 \pm 1.8$ &       0.02 & $(2\,0\,0)5_{0,5} \leftarrow (0\,0\,0)5_{1,4}$ &           1.2~~~~ &      7.6~~~~ & \cite{22DiToScCo}    \\
            $213\,376\,695\,569.3 \pm 1.3$ &       0.02 & $(2\,0\,0)4_{1,4} \leftarrow (0\,0\,0)4_{2,3}$ &         --7.3~~~~ &      8.2~~~~ & \cite{24ToDiCoUb}    \\
            $213\,394\,923\,797.3 \pm 1.4$ &       0.02 & $(2\,0\,0)6_{5,2} \leftarrow (0\,0\,0)7_{4,3}$ &           5.2~~~~ &      2.4~~~~ & \cite{20ToFuSiCs}    \\
            $213\,446\,816\,436.0 \pm 1.6$ &       0.02 & $(2\,0\,0)3_{2,1} \leftarrow (0\,0\,0)3_{3,0}$ &           2.5~~~~ &      2.4~~~~ & \cite{20ToFuSiCs}    \\
            $213\,511\,286\,802.5 \pm 1.4$ &       0.02 & $(2\,0\,0)4_{2,2} \leftarrow (0\,0\,0)4_{3,1}$ &           4.8~~~~ &     12.1~~~~ & \cite{24ToDiCoUb}    \\
            $213\,526\,600\,994.6 \pm 1.6$ &       0.02 & $(2\,0\,0)6_{2,4} \leftarrow (0\,0\,0)6_{3,3}$ &           0.4~~~~ &      5.1~~~~ & \cite{24ToDiCoUb}    \\
            $213\,534\,197\,981.2 \pm 1.5$ &       0.02 & $(2\,0\,0)5_{4,2} \leftarrow (0\,0\,0)6_{3,3}$ &           1.6~~~~ &      7.5~~~~ & \cite{22DiToScCo}    \\
            $213\,539\,473\,234.3 \pm 1.5$ &       0.02 & $(2\,0\,0)3_{2,2} \leftarrow (0\,0\,0)4_{1,3}$ &         --1.1~~~~ &      5.1~~~~ & \cite{24ToDiCoUb}    \\
   $\mathit{213\,663\,124\,477.7 \pm 2.0}$ &       0.01 & $(2\,0\,0)2_{0,2} \leftarrow (0\,0\,0)3_{1,3}$ &           2.8~~~~ &      7.6~~~~ & \cite{24ToDiCoUb}    \\
            $213\,670\,408\,312.7 \pm 1.4$ &       0.02 & $(0\,2\,1)6_{5,1} \leftarrow (0\,0\,0)5_{5,0}$ &        --1300~~~~ &     3000~~~~ & \cite{18MiMoKaKa}    \\
            $213\,731\,119\,700.4 \pm 0.7$ &       0.02 & $(2\,0\,0)6_{3,3} \leftarrow (0\,0\,0)7_{2,6}$ &        --14.6~~~~ &     11.9~~~~ & \cite{24ToDiCoUb}    \\
            $213\,810\,471\,367.1 \pm 1.9$ &       0.02 & $(2\,0\,0)3_{1,3} \leftarrow (0\,0\,0)3_{2,2}$ &           0.9~~~~ &      5.3~~~~ & \cite{20ToFuSiCs}    \\
            $213\,917\,346\,155.5 \pm 1.8$ &       0.02 & $(2\,0\,0)5_{4,1} \leftarrow (0\,0\,0)6_{3,4}$ &           0.3~~~~ &      7.5~~~~ & \cite{22DiToScCo}    \\
   $\mathit{214\,699\,466\,014.1 \pm 1.8}$ &       0.01 & $(2\,0\,0)4_{4,0} \leftarrow (0\,0\,0)5_{3,3}$ &           1.5~~~~ &      3.2~~~~ & \cite{24ToDiCoUb}    \\
            $215\,226\,610\,741.4 \pm 1.7$ &       0.02 & $(2\,0\,0)4_{2,2} \leftarrow (0\,0\,0)5_{1,5}$ &           8.7~~~~ &     10.4~~~~ & \cite{24ToDiCoUb}    \\
            $215\,262\,964\,876.7 \pm 1.7$ &       0.02 & $(2\,0\,0)3_{2,1} \leftarrow (0\,0\,0)4_{1,4}$ &           6.7~~~~ &      8.0~~~~ & \cite{24ToDiCoUb}    \\
            $215\,412\,696\,377.0 \pm 1.2$ &       0.02 & $(2\,0\,0)5_{2,3} \leftarrow (0\,0\,0)6_{1,6}$ &           0.5~~~~ &      7.5~~~~ & \cite{22DiToScCo}    \\
   $\mathit{215\,524\,549\,225.1 \pm 1.7}$ &       0.02 & $(1\,0\,1)5_{2,3} \leftarrow (0\,0\,0)4_{4,0}$ &           2.6~~~~ &      2.6~~~~ & \cite{20ToFuSiCs}    \\
   $\mathit{215\,531\,118\,312.8 \pm 1.9}$ &       0.01 & $(2\,0\,0)2_{2,0} \leftarrow (0\,0\,0)3_{1,3}$ &           3.7~~~~ &      2.4~~~~ & \cite{20ToFuSiCs}    \\
            $215\,854\,572\,219.6 \pm 2.1$ &       0.02 & $(2\,0\,0)5_{1,5} \leftarrow (0\,0\,0)4_{2,2}$ &           4.4~~~~ &      7.7~~~~ & \cite{22DiToScCo}    \\
            $215\,968\,528\,397.7 \pm 1.7$ &       0.02 & $(2\,0\,0)4_{2,3} \leftarrow (0\,0\,0)3_{3,0}$ &      \u{12.4}~~~~ &      4.9~~~~ & \cite{24ToDiCoUb}    \\
            $216\,021\,042\,445.5 \pm 1.6$ &       0.02 & $(2\,0\,0)4_{1,4} \leftarrow (0\,0\,0)3_{2,1}$ &         --7.6~~~~ &      4.9~~~~ & \cite{24ToDiCoUb}    \\
            $216\,269\,473\,019.8 \pm 1.4$ &       0.02 & $(1\,0\,1)4_{3,1} \leftarrow (0\,0\,0)5_{1,4}$ &           0.5~~~~ &      5.3~~~~ & \cite{20ToFuSiCs}    \\
            $216\,320\,509\,854.1 \pm 1.3$ &       0.02 & $(0\,0\,2)4_{1,4} \leftarrow (0\,0\,0)5_{2,3}$ &           500~~~~ &     3000~~~~ & \cite{18MiMoKaKa}    \\
   $\mathit{216\,467\,935\,089.8 \pm 1.9}$ &       0.02 & $(2\,0\,0)4_{2,2} \leftarrow (0\,0\,0)3_{3,1}$ &           7.2~~~~ &      4.9~~~~ & \cite{24ToDiCoUb}    \\
   $\mathit{216\,766\,816\,926.4 \pm 1.9}$ &       0.02 & $(1\,0\,1)5_{2,3} \leftarrow (0\,0\,0)6_{0,6}$ &           4.5~~~~ &      2.4~~~~ & \cite{20ToFuSiCs}    \\
   $\mathit{216\,815\,183\,206.7 \pm 4.6}$ &       0.01 & $(2\,0\,0)2_{0,2} \leftarrow (0\,0\,0)1_{1,1}$ &         --0.8~~~~ &      9.5~~~~ & \cite{24ToDiCoUb}    \\
            $216\,913\,210\,411.2 \pm 1.7$ &       0.02 & $(2\,0\,0)6_{3,4} \leftarrow (0\,0\,0)7_{0,7}$ &         --3.1~~~~ &      3.1~~~~ & \cite{24ToDiCoUb}    \\
   $\mathit{216\,943\,217\,125.0 \pm 1.8}$ &       0.01 & $(2\,0\,0)2_{2,0} \leftarrow (0\,0\,0)2_{1,1}$ &           2.9~~~~ &      5.6~~~~ & \cite{20ToFuSiCs}    \\
            $217\,353\,416\,776.9 \pm 1.4$ &       0.02 & $(2\,0\,0)5_{2,3} \leftarrow (0\,0\,0)4_{3,2}$ &          16.5~~~~ &     11.4~~~~ & \cite{24ToDiCoUb}    \\
            $217\,533\,265\,067.6 \pm 1.5$ &       0.02 & $(2\,0\,0)3_{2,2} \leftarrow (0\,0\,0)3_{1,3}$ &           3.4~~~~ &      5.5~~~~ & \cite{20ToFuSiCs}    \\
   $\mathit{217\,680\,819\,618.1 \pm 1.7}$ &       0.01 & $(1\,0\,1)5_{2,3} \leftarrow (0\,0\,0)5_{2,4}$ &           3.0~~~~ &      6.5~~~~ & \cite{24ToDiCoUb}    \\
            $217\,893\,962\,755.6 \pm 1.5$ &       0.02 & $(2\,0\,0)3_{1,3} \leftarrow (0\,0\,0)2_{0,2}$ &         --0.8~~~~ &      5.2~~~~ & \cite{20ToFuSiCs}    \\
            $218\,089\,528\,336.1 \pm 1.4$ &       0.02 & $(2\,0\,0)5_{2,4} \leftarrow (0\,0\,0)5_{1,5}$ &           8.6~~~~ &      7.9~~~~ & \cite{24ToDiCoUb}    \\
            $218\,300\,740\,560.6 \pm 1.5$ &       0.02 & $(2\,0\,0)4_{4,0} \leftarrow (0\,0\,0)4_{3,1}$ &           1.5~~~~ &      5.2~~~~ & \cite{20ToFuSiCs}    \\
            $218\,430\,303\,226.9 \pm 0.6$ &       0.02 & $(2\,0\,0)6_{2,5} \leftarrow (0\,0\,0)6_{1,6}$ &           3.2~~~~ &      7.5~~~~ & \cite{22DiToScCo}    \\
            $218\,683\,177\,038.4 \pm 1.4$ &       0.02 & $(2\,0\,0)2_{2,0} \leftarrow (0\,0\,0)1_{1,1}$ &           3.5~~~~ &      5.2~~~~ & \cite{20ToFuSiCs}    \\
            $219\,622\,319\,932.4 \pm 1.4$ &       0.02 & $(2\,0\,0)5_{2,4} \leftarrow (0\,0\,0)4_{1,3}$ &           2.4~~~~ &      7.6~~~~ & \cite{22DiToScCo}    \\
            $219\,863\,156\,322.3 \pm 0.7$ &       0.02 & $(2\,0\,0)6_{2,5} \leftarrow (0\,0\,0)5_{1,4}$ &           7.3~~~~ &     11.2~~~~ & \cite{24ToDiCoUb}    \\
            $219\,952\,878\,188.5 \pm 1.4$ &       0.02 & $(0\,0\,2)4_{1,4} \leftarrow (0\,0\,0)5_{0,5}$ &           900~~~~ &     3000~~~~ & \cite{18MiMoKaKa}    \\[2pt]
\end{longtable}
\linespread{\lv}
\normalsize
\endgroup

\begin{tablenotes}
\footnotesize
\item $^a$ New Lamb-dip frequencies, with their standard \citep{nistunc} 
           uncertainties, at pressures displayed in column ``$p$''.
           The nine italicized frequencies were recorded at Amsterdam in 2024, 
           while the other positions were measured at Louvain-la-Neuve in 2025.
\item $^b$ The rovibrational assignments are given as
            $(v'_1\,v'_2\,v'_3)J'_{K'_a,K'_c} 
           \leftarrow (v''_1\,v''_2\,v''_3)J''_{K''_a,K''_c}$
           (see also Sec.~\ref{sec:network}).
\item $^c$ Best literature results, corresponding to previous (Lamb-dip and 
           Doppler-limited) experimental line positions or taken from a SNAPS 
           analysis performed in \cite{24ToDiCoUb}.
           The column ``Dev.'' lists the deviations of the previous frequency
           values from those reported in the first column.
           The column ``Unc.'' contains standard uncertainties provided 
           in the cited publications.
           The deviations, apart from the two underlined cases, remain within
           the claimed \blue{expanded (two-sigma) uncertainties} \citep{nistunc}. 
\end{tablenotes}
\endgroup
\end{widetext}

\begin{widetext}
\begingroup
\linespread{0.9}
\setlength{\tabcolsep}{3.0pt} 

\renewcommand{\ttcap}{}
\renewcommand{\tcap}{Improved frequencies for protected \spec{h216o} lines provided in the
          IAU/CRAF and NAS lists.}
\renewcommand{\tlab}{table:iau16}
\input{H216O_eIAU.tex}
\begin{tablenotes}
\footnotesize
\item $^a$ Approximate (``rest'') frequencies extracted from the primary and 
           supplementary IAU/CRAF reference lists (below 1 THz) or the 
           NAS catalog (above 1 THz).
           The 2631.051 and 2970.801 GHz frequencies, which correspond to the 
           same rovibrational transitions as the 2\,630.960 and 2\,970.800 GHz
           values, respectively, are not included in this table.
\item $^b$ Recommended transition frequencies $\pm$ standard \citep{nistunc} 
           uncertainties, derived 
           in the present work \emph{via} the SNAPS scheme and attached with their 
           rovibrational assignments. 
           Most of these values rely on the new Lamb-dip frequencies given in
           Table~\ref{table:meas}.
\item $^c$ The most accurate laboratory measurements taken from the literature.
           The 6th column (``Dev.'') contains the deviations of the measured 
           positions from the recommended frequencies.
           The 7th column (``Unc.'') includes the standard uncertainties taken 
           from the individual data sources.
           The experimental position of the 2\,640.474 GHz line, reported in 
           \citet{11DrYuPeGu}, does not reflect the zero pressure value, as it 
           was recorded at a pressure higher than 100 Pa, without pressure 
           correction.
\item $^d$ SNAPS estimates utilizing the previous version of the ultraprecise 
           \spec{h216o} network considered in \cite{24UbCsDiCo}.
\item $^e$ The recommended frequencies typeset in italics fully coincide with their 
           direct experimental counterparts, leading to zero deviations in the 
           6th (``Dev.'') column.
           Similarly, where zero deviations are reported in the 4th (``Dev.'') 
           column, the recommended values do not benefit from the newly
           measured Lamb dips.
\end{tablenotes}
\endgroup
\end{widetext}

\begin{widetext}
\linespread{0.9}
\begingroup
\begin{center}
\centering
\hspace{-0.2cm}\includegraphics[width=1.0\linewidth]{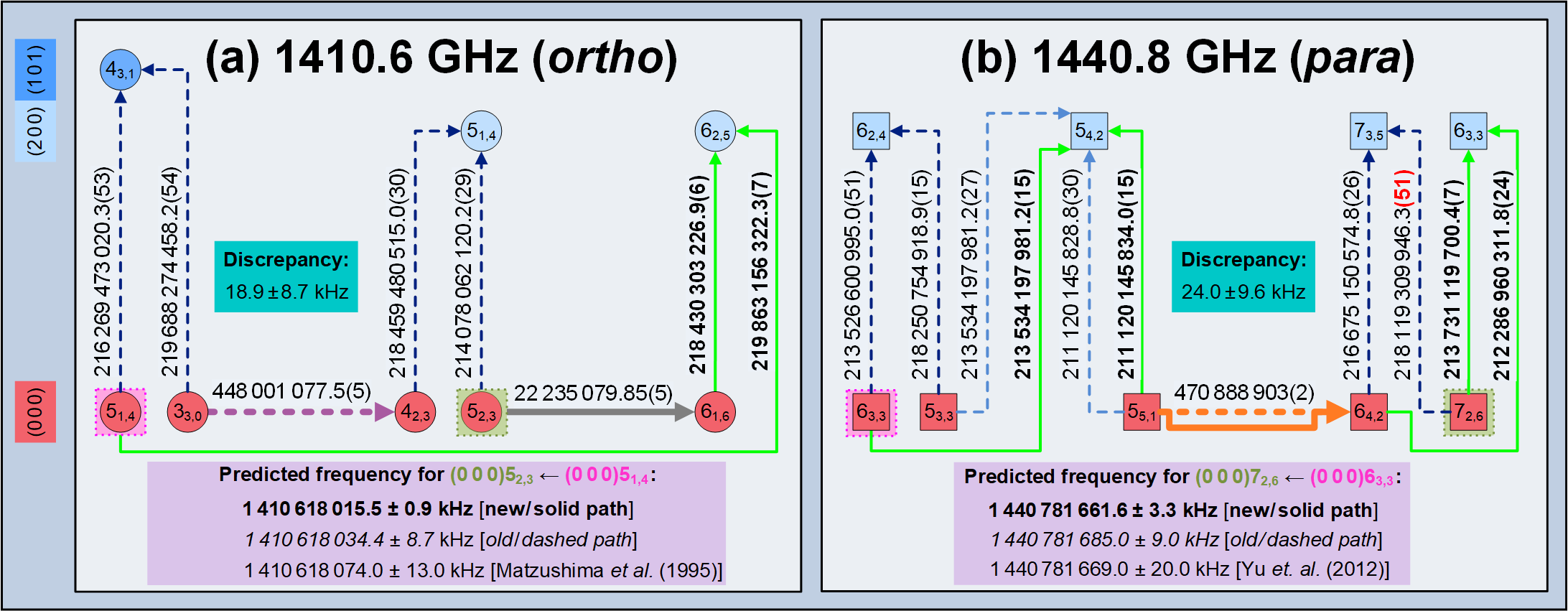}
\vspace{-0.15cm}
\captionof{figure}{\hspace{-0.1cm}Comparison of new and previous 
         kHz-accuracy paths for two 
         selected radio lines of Table~\ref{table:iau16}.
         \normalfont{The \emph{ortho} and \emph{para} states of this figure 
         are designated with circles and squares, respectively.
         For these energy levels, the $J_{K_a,K_c}$ labels are written out 
         explicitly, while the $(v_1\,v_2\,v_3)$ triplets are shown in the 
         left-side color legend.
         The green arrows denote new Lamb-dip transitions, while those in 
         dark blue, gray, orange, purple, and light blue are extremely 
         accurate lines collected from \cite{20ToFuSiCs}, \cite{69Kukolich}, 
         \cite{06GoMaGuKn}, \cite{09CaPuHaGa}, and \cite{24UbCsDiCo}, respectively.
         For the pure rotational transitions included on the paths, the arrows 
         are thickened.
         The numbers on the arrows are frequencies in kHz, with the standard
         \citep{nistunc} uncertainties of the last digits in parentheses.
         The solid arrows constitute the new (smallest-uncertainty) paths 
         between their starting and ending states, surrounded by dotted 
         magenta and green boxes, respectively.
         The red uncertainty value was slightly underestimated in
         \citet{20ToFuSiCs}, the true uncertainty must be around 15 kHz. 
         The dashed arrows represent the previous paths, forming cycles with the 
         solid ones.
         The orange line, illustrated as a double arrow, is part of both paths.
         The rest frequencies, related to lines between the starting and 
         ending states of the two paths, are placed at the top of both panels.
         The light mauve and teal blue boxes provide predicted frequencies and 
         discrepancies, along with their standard uncertainties, 
         respectively.
         The direct experimental values are also listed in the mauve boxes.
         These two examples correspond to the highest improvements achieved for
         the protected \emph{ortho}- and \emph{para}-\spec{h216o} transitions.
         }
        \label{fig:iau_snaps}
} 
\end{center}
\endgroup
\end{widetext}

The list of \const{55} lines, measured during this study for \spec{h216o}, is 
presented in Table~\ref{table:meas}, where the nine frequencies recorded in 
Amsterdam are typeset in italics.
The typical accuracy of the new measurements is around 2 kHz, thereby 
improving upon our previous experimental \citep{20ToFuSiCs,22DiToScCo} or 
SNAPS-predicted \citep{24ToDiCoUb} line-center positions.
For eight transitions, only Doppler-broadened observations were previously 
available, with a claimed uncertainty of 3 MHz \citep{18MiMoKaKa}.

\section{Network-based frequency predictions} \label{sec:PF}
The ultraprecise \spec{h216o} network, augmented with the newly measured lines 
presented in Table~\ref{table:meas}, was used to extract highly accurate 
frequencies for radio lines.
Figure~\ref{fig:iau_snaps} illustrates how the predicted frequencies have been 
extended from this network, applying the Ritz principle \citep{08Ritz} in a 
successive fashion.

In these examples, a connection has to be made between opposite-parity levels in 
the ground vibrational state and an odd number of known rotational 
lines must be included.
The first nine IAU/CRAF transitions, see Table~\ref{table:iau16}, which were 
measured at an accuracy better than 1 kHz in microwave studies \citep{69Kukolich,
06GoMaGuKn,09CaPuHaGa}, serve this purpose.
This extreme precision could be achieved only for one SNAPS prediction, see
Fig.~\ref{fig:iau_snaps}(a), relying on two of the four sub-kHz accuracy lines
reported in Table~\ref{table:meas}.

For the remaining \const{51} lines in Table~\ref{table:iau16}, all part of the 
IAU/CRAF and NAS lists, the frequencies could be improved, on average,
by a factor of 3,
with respect to our previous SNAPS analysis \citep{24ToDiCoUb}, and even better 
when compared to the direct terahertz measurements \citep{95MaOdIwTs,98ChPaEv,
11DrYuPeGu,12YuPeDrMa}.
The new network-assisted measurements made it possible to utilize the sub-kHz 
accuracy near-infrared transition of \cite{22KaLaChCa}, yielding a frequency 
uncertainty of 2.6 kHz for the 1163 GHz line.
For a comparison of the various frequency estimates available 
for the protected \spec{h216o} lines \citep{46ToMe,48GoWeHiSt,51Jen,54KiGo,69Kukolich,
70StSt,71Huiszoon,71StBe,72FlCaVa,72LuHeCoGo,76FlGi,78KaKaKy,80KaKy,81Partridg,
83BuFeKaPo,83HeMeLu,83MeLuHe,85Johns,86GuRa,89BaGoKhHa,95MaKr,95MaOdIwTs,96BrPl,
98BaGoCaGa,98ChPaEv,04CoPiVeLa,05Golubiat,06GoMaGuKn,07KoTrGoPa,09CaPuHaGa,11DrYuPeGu,
11Koshelev,12YuPeDrMa,13TrKoViPa,13YuPeDr,14CoMaPi,20MiBeOdTr,20ToFuSiCs,22DiToScCo,
22ToKoMiPi,24KaMiKoCa,24MiKaKoCa}, see the Supplemental Information.

It should be stressed that all transition frequencies included in the 
ultraprecise \spec{h216o} network relate to zero-pressure values, 
either through 
($i$) measurements at very low (0.01\,--\,0.02 Pa) pressures,
($ii$) explicit extrapolation to zero pressure, or 
($iii$) including the possible pressure-shift effects in the uncertainty budget.
Hence, the listed values for the protected radio lines also pertain to 
zero-pressure values, with special relevance for the extremely low pressures in 
interstellar space.

\section{Conclusion}

In the present study, an ultraprecise spectroscopic network, corresponding to 
quantum states and transitions of the \spec{h216o} water isotopologue has been 
extended and refined by the precise frequency calibration of \const{55} 
rovibrational transitions determined using the NICE-OHMS measurement technology.
Of the \const{60} radio lines protected by the International Astronomical Union 
and the Panel on Frequency Allocations of the US National Academy of Sciences, 
51 have been considerably improved in accuracy due to our present and previous 
Lamb-dip positions.
Specifically, their uncertainties, falling into the 10\,--\,200 kHz range, have been 
brought to below \const{7.5} kHz.
For the nine lowest-frequency transitions, all known with an accuracy better than 1 
kHz, no improvements could be achieved. 
These lines play an essential role within the network, because they connect the 
\emph{even}- and \emph{odd}-parity levels of the ground vibrational state.

In principle, a similar analysis might be conducted for \spec{h218o}, as well. 
A large database of Lamb-dip lines has already been collected \citep{21DiToSiCo}, and 
a partial procedure for extracting radio lines from this database has been executed 
previously~\citep{24UbCsDiCo}. 
For the \spec{hdo} isotopologue, 
such a procedure was halted in view of parity-pair-mixing effects that limited
the accuracy of the Lamb dips measured in \citet{22DiToCoSa}.

\begin{center} SUPPLEMENTAL INFORMATION \end{center}\vspace{-0.0cm}
Additional data related to this work are available at 
\url{https://zenodo.org/records/16757281}.
Table S1 contains a list of new experimental \spec{h216o} rovibrational transitions 
recorded for the present paper with the NICE-OHMS technique at Amsterdam and 
Louvain-la-Neuve.
Table S2 provides the paths within the ultraprecise \spec{h216o} network for all 
radio lines given in Table~\ref{table:iau16}, the best paths defining radio
frequencies\blue{, as well as the collection of data sources used in the comparison.
Table S3 includes basic cycles associated with the shortest-path-based forest of 
the ultraprecise \spec{h216o} spectroscopic network.}

\begin{center} ACKNOWLEDGEMENTS \end{center}\vspace{-0.0cm}
This work was funded by LASER\-LAB-EU\-ROPE (grant no. 654148), a European Union's 
Horizon 2020 research and innovation programme.
The present study is part of a European Partnership on a Metrology project, which is 
co-financed by the European Unionâ€™s Horizon Europe Research and Innovation Programme 
and by the Participating States (funder ID: 10.13039/100019599, grant no.: 23FUN04 
COMOMET).
The research performed in Budapest received support from the National Research,
Development and Innovation Office of Hungary (NKFIH, grants K138233 and PD145972).
A.A., A. B. and C. L.  acknowledge the FNRS CDR (grant
no. J.0113.23) and MIS (grant no. F.4038.24) for financial support.

\begingroup
\scriptsize
\singlespacing

\bibliographystyle{aasjournal}

\begin{thebibliography}{}
\expandafter\ifx\csname natexlab\endcsname\relax\def\natexlab#1{#1}\fi
\providecommand{\url}[1]{\href{#1}{#1}}
\providecommand{\dodoi}[1]{doi:~\href{http://doi.org/#1}{\nolinkurl{#1}}}
\providecommand{\doeprint}[1]{\href{http://ascl.net/#1}{\nolinkurl{http://ascl.net/#1}}}
\providecommand{\doarXiv}[1]{\href{https://arxiv.org/abs/#1}{\nolinkurl{https://arxiv.org/abs/#1}}}

\bibitem[{Bauer {et~al.}(1998)Bauer, Godon, Carlier, \& Gamache}]{98BaGoCaGa}
Bauer, A., Godon, M., Carlier, J., \& Gamache, R. 1998, J. Quant. Spectrosc. Rad. Transf., 59, 273

\bibitem[{Bauer {et~al.}(1989)Bauer, Godon, Kheddar, \& Hartmann}]{89BaGoKhHa}
Bauer, A., Godon, M., Kheddar, M., \& Hartmann, J. 1989, J. Quant. Spectrosc. Rad. Transf., 41, 49

\bibitem[{Brown \& Plymate(1996)}]{96BrPl}
Brown, L.~R., \& Plymate, C. 1996, J. Quant. Spectrosc. Rad. Transf., 56, 263

\bibitem[{Burenin {et~al.}(1983)Burenin, Fevralskikh, Karyakin, Polyansky, \& Shapin}]{83BuFeKaPo}
Burenin, A.~V., Fevralskikh, T.~M., Karyakin, E.~N., Polyansky, O.~L., \& Shapin, S.~M. 1983, J. Mol. Spectrosc., 100, 182

\bibitem[{Cazzoli {et~al.}(2009)Cazzoli, Puzzarini, Harding, \& Gauss}]{09CaPuHaGa}
Cazzoli, G., Puzzarini, C., Harding, M.~E., \& Gauss, J. 2009, Chem. Phys. Lett., 473, 21

\bibitem[{Cazzoli {et~al.}(2004)Cazzoli, Puzzarini, \& Lapinov}]{04CaPuLa}
Cazzoli, G., Puzzarini, C., \& Lapinov, A. 2004, Astrophys J., 611, 615

\bibitem[{Cernicharo {et~al.}(2006)Cernicharo, Goicoechea, Daniel, Lerate, Barlow, Swinyard, van Dishoeck, Lim, Viti, \& Yates}]{06CeGoDaLe}
Cernicharo, J., Goicoechea, J.~R., Daniel, F., {et~al.} 2006, Astrophys. J, 649, L33

\bibitem[{Chance {et~al.}(1998)Chance, Park, \& Evenson}]{98ChPaEv}
Chance, K.~V., Park, K., \& Evenson, K.~M. 1998, J. Quant. Spectrosc. Rad. Transf., 59, 687

\bibitem[{Chen {et~al.}(2018)Chen, Hua, Tao, Sun, Liu, \& Hu}]{18ChHuTaSu}
Chen, J., Hua, T.-P., Tao, L.-G., {et~al.} 2018, J. Quant. Spectrosc. Rad. Transf., 205, 91

\bibitem[{Cheung {et~al.}(1969)Cheung, Rank, Townes, Thornton, \& Welch}]{69ChRaToTh}
Cheung, A.~C., Rank, D.~M., Townes, C.~H., Thornton, D.~D., \& Welch, W.~J. 1969, Nature, 221, 626â€“628

\bibitem[{{Committee on Radio Frequencies Board on Physics and Astronomy, National Academies of Sciences -- Engineering -- Medicine}(2015)}]{15Committe}
{Committee on Radio Frequencies Board on Physics and Astronomy, National Academies of Sciences -- Engineering -- Medicine}. 2015, {Handbook of Frequency Allocations and Spectrum Protection for Scientific Uses} (National Academy of Sciences, Washington, USA)

\bibitem[{Coudert {et~al.}(2014)Coudert, Martin-Drumel, \& Pirali}]{14CoMaPi}
Coudert, L.~H., Martin-Drumel, M.-A., \& Pirali, O. 2014, J. Mol. Spectrosc., 303, 36

\bibitem[{Coudert {et~al.}(2004)Coudert, Pirali, Vervloet, Lanquetin, \& Camy-Peyret}]{04CoPiVeLa}
Coudert, L.~H., Pirali, O., Vervloet, M., Lanquetin, R., \& Camy-Peyret, C. 2004, J. Mol. Spectrosc., 228, 471

\bibitem[{Cozijn {et~al.}(2018)Cozijn, Dupr\'e, Salumbides, Eikema, \& Ubachs}]{18CoDuSaEi}
Cozijn, F. M.~J., Dupr\'e, P., Salumbides, E.~J., Eikema, K. S.~E., \& Ubachs, W. 2018, Phys. Rev. Lett., 120, 153002

\bibitem[{Cs\'asz\'ar \& Furtenbacher(2011)}]{11CsFu}
Cs\'asz\'ar, A.~G., \& Furtenbacher, T. 2011, J. Mol. Spectrosc., 266, 99

\bibitem[{de~Lucia {et~al.}(1972)de~Lucia, Helminger, Cook, \& Gordy}]{72LuHeCoGo}
de~Lucia, F.~C., Helminger, P., Cook, R.~L., \& Gordy, W. 1972, Phys. Rev. A, 5, 487

\bibitem[{Diouf {et~al.}(2022{\natexlab{a}})Diouf, T\'obi\'as, Cozijn, Salumbides, F\'abri, Puzzarini, Cs\'asz\'ar, \& Ubachs}]{22DiToCoSa}
Diouf, M.~L., T\'obi\'as, R., Cozijn, F. M.~J., {et~al.} 2022{\natexlab{a}}, Opt. Express, 30, 46040

\bibitem[{Diouf {et~al.}(2021)Diouf, T\'obi\'as, Simk\'o, Cozijn, Salumbides, Ubachs, \& Cs\'asz\'ar}]{21DiToSiCo}
Diouf, M.~L., T\'obi\'as, R., Simk\'o, I., {et~al.} 2021, J. Phys. Chem. Ref. Data, 50, 023106

\bibitem[{Diouf {et~al.}(2022{\natexlab{b}})Diouf, T\'obi\'as, {van der Schaaf}, Cozijn, Salumbides, Cs\'asz\'ar, \& Ubachs}]{22DiToScCo}
Diouf, M.~L., T\'obi\'as, R., {van der Schaaf}, T.~S., {et~al.} 2022{\natexlab{b}}, Mol. Phys., 120, e2050430

\bibitem[{Drouin {et~al.}(2011)Drouin, Yu, Pearson, \& Gupta}]{11DrYuPeGu}
Drouin, B.~J., Yu, S., Pearson, J.~C., \& Gupta, H. 2011, J. Mol. Struct., 1066, 2

\bibitem[{Endres {et~al.}(2016)Endres, Schlemmer, Schilke, Stutzki, \& MĂĽller}]{16ChScScSt}
Endres, C.~P., Schlemmer, S., Schilke, P., Stutzki, J., \& MĂĽller, H. S.~P. 2016, J. Mol. Spectrosc., 327, 95

\bibitem[{Flaud {et~al.}(1972)Flaud, Camy-Peyret, \& Valentin}]{72FlCaVa}
Flaud, J.-M., Camy-Peyret, C., \& Valentin, A. 1972, J. Phys. (Paris), 33, 741

\bibitem[{Fleming \& Gibson(1976)}]{76FlGi}
Fleming, J.~W., \& Gibson, M.~J. 1976, J. Mol. Spectrosc., 62, 326

\bibitem[{Furtenbacher {et~al.}(2020{\natexlab{a}})Furtenbacher, T\'obi\'as, Tennyson, Polyansky, \& Cs\'asz\'ar}]{20FuToTePo}
Furtenbacher, T., T\'obi\'as, R., Tennyson, J., Polyansky, O.~L., \& Cs\'asz\'ar, A.~G. 2020{\natexlab{a}}, J. Phys. Chem. Ref. Data, 49, 033101

\bibitem[{Furtenbacher {et~al.}(2020{\natexlab{b}})Furtenbacher, T\'obi\'as, Tennyson, Polyansky, Kyuberis, Ovsyannikov, Zobov, \& Cs\'asz\'ar}]{20FuToTePob}
Furtenbacher, T., T\'obi\'as, R., Tennyson, J., {et~al.} 2020{\natexlab{b}}, J. Phys. Chem. Ref. Data, 49, 043103

\bibitem[{Gianfrani {et~al.}(2024)Gianfrani, Hu, \& Ubachs}]{24GiHuUb}
Gianfrani, L., Hu, S.-M., \& Ubachs, W. 2024, La Rivista del Nuovo Cimento, 47, 229â€“298

\bibitem[{Golden {et~al.}(1948)Golden, Wentink, Hillger, \& Strandberg}]{48GoWeHiSt}
Golden, S., Wentink, T., Hillger, R., \& Strandberg, M. W.~P. 1948, Phys. Rev., 73, 92

\bibitem[{Golubiatnikov(2005)}]{05Golubiat}
Golubiatnikov, G.~Y. 2005, J. Mol. Spectrosc., 230, 196

\bibitem[{Golubiatnikov {et~al.}(2006)Golubiatnikov, Markov, Guarnieri, \& Knochel}]{06GoMaGuKn}
Golubiatnikov, G.~Y., Markov, V.~N., Guarnieri, A., \& Knochel, R. 2006, J. Mol. Spectrosc., 240, 251

\bibitem[{Guelachvili \& Rao(1986)}]{86GuRa}
Guelachvili, G., \& Rao, K.~N. 1986, {Handbook of Infrared Standards} (Academic Press, Orlando)

\bibitem[{Helminger {et~al.}(1983)Helminger, Messer, \& de~Lucia}]{83HeMeLu}
Helminger, P., Messer, J.~K., \& de~Lucia, F.~C. 1983, Appl. Phys. Lett., 42, 309

\bibitem[{Huiszoon(1971)}]{71Huiszoon}
Huiszoon, C. 1971, Rev. Sci. Instrum., 42, 477

\bibitem[{Jen(1951)}]{51Jen}
Jen, C.~K. 1951, Phys. Rev., 81, 197

\bibitem[{Johns(1985)}]{85Johns}
Johns, J. W.~C. 1985, J. Opt. Soc. Am. B, 2, 1340

\bibitem[{Karlovets {et~al.}(2024)Karlovets, Mikhailenko, Koroleva, \& Campargue}]{24KaMiKoCa}
Karlovets, E.~V., Mikhailenko, S.~N., Koroleva, A.~O., \& Campargue, A. 2024, J. Quant. Spectrosc. Rad. Transf., 314, 108829

\bibitem[{Kassi {et~al.}(2022)Kassi, Lauzin, Chaillot, \& Campargue}]{22KaLaChCa}
Kassi, S., Lauzin, C., Chaillot, J., \& Campargue, A. 2022, Phys. Chem. Chem. Phys., 24, 23164

\bibitem[{Kassi {et~al.}(2018)Kassi, Stoltmann, Casado, Daeron, \& Campargue}]{18KaStCaDa}
Kassi, S., Stoltmann, T., Casado, M., Daeron, M., \& Campargue, A. 2018, J. Chem. Phys., 148, 054201

\bibitem[{Kauppinen {et~al.}(1978)Kauppinen, K\"arkk\"ainen, \& Kyr\"{o}}]{78KaKaKy}
Kauppinen, J., K\"arkk\"ainen, T., \& Kyr\"{o}, E. 1978, J. Mol. Spectrosc., 71, 15

\bibitem[{Kauppinen \& Kyr\"{o}(1980)}]{80KaKy}
Kauppinen, J., \& Kyr\"{o}, E. 1980, J. Mol. Spectrosc., 84, 405

\bibitem[{King \& Gordy(1954)}]{54KiGo}
King, W.~C., \& Gordy, W. 1954, Phys. Rev., 93, 407

\bibitem[{Koshelev(2011)}]{11Koshelev}
Koshelev, M.~A. 2011, J. Quant. Spectrosc. Rad. Transf., 112, 550

\bibitem[{Koshelev {et~al.}(2007)Koshelev, Tretyakov, Golubiatnikov, Parshin, Markov, \& Koval}]{07KoTrGoPa}
Koshelev, M.~A., Tretyakov, M.~Y., Golubiatnikov, G.~Y., {et~al.} 2007, J. Mol. Spectrosc., 241, 101

\bibitem[{Kukolich(1969)}]{69Kukolich}
Kukolich, S.~G. 1969, J. Chem. Phys., 50, 3751

\bibitem[{Markov \& Krupnov(1995)}]{95MaKr}
Markov, V.~N., \& Krupnov, A.~F. 1995, J. Mol. Spectrosc., 172, 211

\bibitem[{Matsushima {et~al.}(1995)Matsushima, Odashima, Iwasaki, Tsunekawa, \& Takagi}]{95MaOdIwTs}
Matsushima, F., Odashima, H., Iwasaki, T., Tsunekawa, S., \& Takagi, K. 1995, J. Mol. Spectrosc., 352, 371

\bibitem[{Messer {et~al.}(1983)Messer, de~Lucia, \& Helminger}]{83MeLuHe}
Messer, J.~K., de~Lucia, F.~C., \& Helminger, P. 1983, Int. J Infrared Milli., 4, 505

\bibitem[{Miani \& Tennyson(2004)}]{04MiTe}
Miani, A., \& Tennyson, J. 2004, J. Chem. Phys., 120, 2732

\bibitem[{Mikhailenko {et~al.}(2020)Mikhailenko, B\'eguier, Odintsova, Tretyakov, Pirali, \& Campargue}]{20MiBeOdTr}
Mikhailenko, S.~N., B\'eguier, S., Odintsova, T.~A., {et~al.} 2020, J. Quant. Spectrosc. Rad. Transf., 253, 107105

\bibitem[{Mikhailenko {et~al.}(2024)Mikhailenko, Karlovets, Koroleva, \& Campargue}]{24MiKaKoCa}
Mikhailenko, S.~N., Karlovets, E.~V., Koroleva, A.~O., \& Campargue, A. 2024, J. Phys. Chem. Ref. Data, 53

\bibitem[{Mikhailenko {et~al.}(2018)Mikhailenko, Mondelain, Karlovets, Kassi, \& Campargue}]{18MiMoKaKa}
Mikhailenko, S.~N., Mondelain, D., Karlovets, E.~V., Kassi, S., \& Campargue, A. 2018, J. Quant. Spectrosc. Rad. Transf., 206, 163

\bibitem[{M\"uller {et~al.}(2005)M\"uller, Schl\"oder, Stutzki, \& Winnewisser}]{05MuScStWi}
M\"uller, H. S.~P., Schl\"oder, F., Stutzki, J., \& Winnewisser, G. 2005, J. Mol. Struct., 742, 215

\bibitem[{{NIST/SEMATECH}(2012)}]{nistunc}
{NIST/SEMATECH}. 2012, {NIST/SEMATECH e-Handbook of Statistical Methods, Measurement Process Characterization, Section 2.5.7}, \url{https://www.itl.nist.gov/div898/handbook/mpc/section5/mpc57.htm}

\bibitem[{Partridge(1981)}]{81Partridg}
Partridge, R.~H. 1981, J. Mol. Spectrosc., 87, 429

\bibitem[{Pickett {et~al.}(1998)Pickett, Poynter, Cohen, Delitsky, Pearson, \& M\"uller}]{98PiPoCoDe}
Pickett, H., Poynter, R., Cohen, E., {et~al.} 1998, J. Quant. Spectrosc. Rad. Transf., 60, 883

\bibitem[{Ritz(1908)}]{08Ritz}
Ritz, W. 1908, Astrophys. J., 28, 237

\bibitem[{Solomon \& Klemperer(1972)}]{72SoKl}
Solomon, P., \& Klemperer, W. 1972, Astrophys. J., 178, 389

\bibitem[{Steenbeckeliers \& Bellet(1971)}]{71StBe}
Steenbeckeliers, G., \& Bellet, J. 1971, C. R. Acad. Sci. B Phys., 273, 471

\bibitem[{Stephenson \& Strauch(1970)}]{70StSt}
Stephenson, D.~A., \& Strauch, R.~G. 1970, J. Mol. Spectrosc., 35, 494

\bibitem[{Supplee {et~al.}(1994)Supplee, Whittaker, \& Lenth}]{Supplee1994}
Supplee, J.~M., Whittaker, E.~A., \& Lenth, W. 1994, Appl. Opt., 33, 6294

\bibitem[{Tielens \& Hagen(1982)}]{82TiHa}
Tielens, A., \& Hagen, W. 1982, Astron. Astrophys., 114, 245

\bibitem[{T\'obi\'as {et~al.}(2024)T\'obi\'as, Diouf, Cozijn, Ubachs, \& Cs\'asz\'ar}]{24ToDiCoUb}
T\'obi\'as, R., Diouf, M.~L., Cozijn, F.~M., Ubachs, W., \& Cs\'asz\'ar, A.~G. 2024, Commun. Chem., 7, 34

\bibitem[{T\'obi\'as {et~al.}(2020)T\'obi\'as, Furtenbacher, Simk\'o, Cs\'asz\'ar, Diouf, Cozijn, Staa, Salumbides, \& Ubachs}]{20ToFuSiCs}
T\'obi\'as, R., Furtenbacher, T., Simk\'o, I., {et~al.} 2020, Nat. Commun., 11, 1708

\bibitem[{Toureille {et~al.}(2022)Toureille, Koroleva, Mikhailenko, Pirali, \& Campargue}]{22ToKoMiPi}
Toureille, M., Koroleva, A.~O., Mikhailenko, S.~N., Pirali, O., \& Campargue, A. 2022, J. Quant. Spectrosc. Rad. Transf., 291, 108326

\bibitem[{Townes \& Merritt(1946)}]{46ToMe}
Townes, C.~H., \& Merritt, F.~R. 1946, Phys. Rev., 70, 558

\bibitem[{Tretyakov {et~al.}(2013)Tretyakov, Koshelev, Vilkov, Parshin, \& Serov}]{13TrKoViPa}
Tretyakov, M.~Y., Koshelev, M.~A., Vilkov, I.~N., Parshin, V.~V., \& Serov, E.~A. 2013, J. Quant. Spectrosc. Rad. Transf., 114, 109

\bibitem[{Ubachs {et~al.}(2024)Ubachs, Cs\'asz\'ar, Diouf, Cozijn, \& T\'obi\'as}]{24UbCsDiCo}
Ubachs, W., Cs\'asz\'ar, A.~G., Diouf, M.~L., Cozijn, F.~M., \& T\'obi\'as, R. 2024, ACS Earth Space Chem., 8, 1901

\bibitem[{van Dishoeck {et~al.}(2013)van Dishoeck, Herbst, \& Neufeld}]{13DiHeNe}
van Dishoeck, E.~F., Herbst, E., \& Neufeld, D.~A. 2013, Chem. Rev., 113, 9043

\bibitem[{WeiĂź {et~al.}(2013)WeiĂź, De~Breuck, Marrone, Vieira, Aguirre, Aird, Aravena, Ashby, Bayliss, Benson, BĂ©thermin, Biggs, Bleem, Bock, Bothwell, Bradford, Brodwin, Carlstrom, Chang, Chapman, Crawford, Crites, de~Haan, Dobbs, Downes, Fassnacht, George, Gladders, Gonzalez, Greve, Halverson, Hezaveh, High, Holder, Holzapfel, Hoover, Hrubes, Husband, Keisler, Lee, Leitch, Lueker, Luong-Van, Malkan, McIntyre, McMahon, Mehl, Menten, Meyer, Murphy, Padin, Plagge, Reichardt, Rest, Rosenman, Ruel, Ruhl, Schaffer, Shirokoff, Spilker, Stalder, Staniszewski, Stark, Story, Vanderlinde, Welikala, \& Williamson}]{13WeBrMaVi}
WeiĂź, A., De~Breuck, C., Marrone, D.~P., {et~al.} 2013, Astrophys. J., 767, 88

\bibitem[{Wright {et~al.}(2000)Wright, van Dishoeck, Black, Feuchtgruber, Cernicharo, Gonzalez-Alfonso, \& deGraauw}]{00WrDiBlFe}
Wright, C.~M., van Dishoeck, E.~F., Black, J.~H., {et~al.} 2000, Astron. Astrophys., 358, 89

\bibitem[{Yu {et~al.}(2013)Yu, Pearson, \& Drouin}]{13YuPeDr}
Yu, S., Pearson, J.~C., \& Drouin, B.~J. 2013, J. Mol. Spectrosc., 288, 7

\bibitem[{Yu {et~al.}(2012)Yu, Pearson, Drouin, Martin-Drumel, Pirali, Vervloet, Coudert, M\"uller, \& Br\"unken}]{12YuPeDrMa}
Yu, S., Pearson, J.~C., Drouin, B.~J., {et~al.} 2012, J. Mol. Spectrosc., 279, 16

\end{thebibliography}
\endgroup

\end{document}